\documentclass[3p,times,twocolumn]{elsarticle}

\usepackage{ecrc}

\volume{00}

\firstpage{1}

\journalname{Nuclear Physics B Proceedings Supplement}

\runauth{A. Vicente}

\jid{nuphbp}

\jnltitlelogo{Nuclear Physics B Proceedings Supplement}

\usepackage{amsmath,amsfonts,amssymb,amsthm}
\usepackage[figuresright]{rotating}

\begin{document}

\begin{frontmatter}

\dochead{}

\title{Charged lepton flavor violation beyond minimal supersymmetry}

\author{A. Vicente}

\ead{Avelino.Vicente@ulg.ac.be}

\address{IFPA, Dep. AGO, Universit\'e de Li\`ege, Bat B5, Sart-Tilman B-4000 Li\`ege 1, Belgium}

\begin{abstract}
We discuss charged lepton flavor violation in supersymmetric models
with extended leptonic sectors at low energies. Contrary to the usual
high-scale realizations of the seesaw mechanism, these non-minimal
supersymmetric models have new superfields and/or interactions at the
supersymmetric scale. As a consequence of this, the resulting lepton
flavor violating phenomenology may be very different from that of
minimal models.
\end{abstract}


\end{frontmatter}

\section{Introduction}
\label{sec:intro}

Since the discovery of neutrino oscillations, the violation of lepton
flavor is a well-established fact. However this behavior has never
been observed in processes involving charged leptons. This is well
understood in the Standard Model (SM) minimally extended to include
neutrino masses. Since the only source of lepton flavor violation
(LFV) is the neutrino mass matrix itself, all LFV processes are highly
suppressed. This makes the observation of charged lepton flavor
violation a clear sign of new physics.

New sources of LFV can be found in most extensions of the leptonic
sector. These may be caused by new interactions (renormalizable or
non-renormalizable) or by entire new sectors coupled to the charged
leptons. In the case of supersymmetry (SUSY), the new degrees of
freedom provided by the superpartners of the SM leptons have
potentially large contributions to LFV processes, leading to
stringent constraints on the slepton masses and mixings.

Most papers on SUSY phenomenology focus on minimal models, such as the
Minimal Supersymmetric Standard Model (MSSM)\footnote{For a recent
  work on LFV in the MSSM see \cite{Arana-Catania:2013nha}.}. However,
neutrinos are massless in the MSSM and thus the leptonic sector must
be further extended in order to account for the observed neutrino
masses and mixings. In general, this can be done in two different
ways:

\begin{itemize}

\item High-energy extensions: In this family of models the new degrees
  of freedom responsible for neutrino masses live at very high energy
  scales. Therefore, the physics at the SUSY scale is well described
  by the MSSM and the effect of the heavy fields is induced by
  Renormalization Group Equations running.

\item Low-energy extensions: In this family of models the new degrees
  of freedom are present at low energies. These include new particles
  and/or interactions. Their effect on the phenomenology is direct, as
  they fully participate at the SUSY scale.

\end{itemize}

The first category can be seen as \emph{Minimal SUSY}, since the
low-energy theory is just the MSSM. These are the most studied SUSY
neutrino mass models in the literature and the simplest example is the
classical SUSY seesaw\footnote{By \emph{classical SUSY seesaw} we mean
  the usual realization of the seesaw mechanism where the seesaw
  messengers have masses close to the unification scale.}. On
the contrary, models belonging to the second category can be seen as
\emph{non-minimal SUSY}. Here we will concentrate on this case and
discuss some recent works related to non-minimal extensions of the
leptonic sector in SUSY models. As will be clear from the following
examples, non-minimal SUSY models may have a very different LFV
phenomenology from that of minimal ones.

\section{R-parity violating models}
\label{sec:RPV}

\begin{figure*}
\begin{center}
\vspace{5mm}
\includegraphics[width=0.38\textwidth]{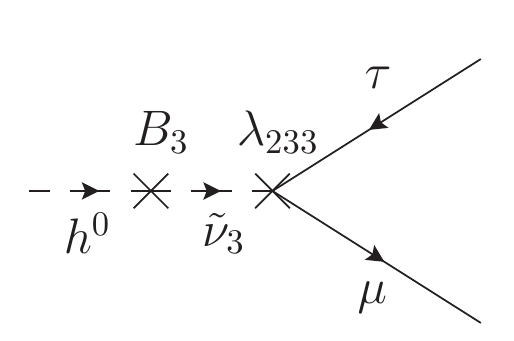}
\hspace*{5mm}
\includegraphics[width=0.38\textwidth]{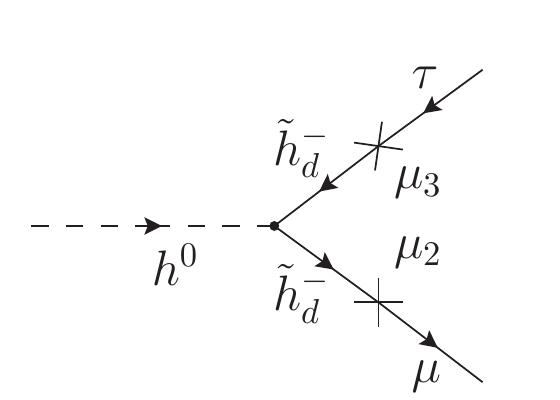}
\end{center}
\caption{Tree-level R-parity violating contributions to $h \to \mu
  \tau$. On the left, a $B \, \lambda$ contribution. On the right, a
  $\mu_i^2$ contribution. Figure borrowed from \cite{Arhrib:2012mg}.}
\label{fig:contributions}
\end{figure*}

R-parity violation (RPV) is usually regarded as a dangerous
possibility due to the non-observation of proton decay. For that
reason, most SUSY models simply assume this \emph{ad-hoc} symmetry
that forbids all renormalizable baryon ($B$) and lepton ($L$) number
violating operators. However, there is no need to forbid all these
operators. In fact, in order to stabilize the proton it is sufficient
to get rid of just one of these terms, namely the baryon number
violating one.

Furthermore, the presence of R-parity violating couplings leads to a
much richer collider phenomenology due to the decay of the lightest
supersymmetric particle (LSP). This adds a further step in the SUSY
decay chains at the LHC and changes dramatically the resulting
signatures \cite{Dreiner:2012wm,Dreiner:2012ec}. This has also been
recently considered in order to relax the stringent bounds on the
squark and gluino masses, otherwise increased to values beyond
expectations based on naturalness arguments, see for example
\cite{Graham:2012th,Hanussek:2012eh}.

In what concerns the leptonic sector, the lepton number violating
interactions induce Majorana neutrino masses
\cite{Hall:1983id,Ross:1984yg}. These models are thus a very
economical framework to explain neutrino masses and mixings and, as we
will discuss below, usually have many non-standard LFV signatures.

Finally, the tight connection between neutrino masses and the decay of
the LSP allows us to have a direct probe for a general class of these
models (those where the relevant L violating operator is the bilinear
term $\mu_i \widehat L_i \widehat H_u$) at the LHC, see the recent
Ref. \cite{deCampos:2012pf}.

\subsection{Higgs boson LFV decays}
\label{subsec:hmutau}

After the recent discovery of the long-awaited Higgs boson, the focus
has shifted towards the experimental determination of its
properties. On the one hand, they are crucial to conclude that the
particle discovered at the LHC actually corresponds to the Higgs
boson, the footprint left by the spontaneous breaking of the SM gauge
symmetry. On the other hand, even if this new particle turns out be a
very SM-like Higgs boson, there is still hope for some non-standard
properties.

The LFV decay $h \to \ell_i \ell_j$, with $i \ne j$, has recently got
some attention \cite{Harnik:2012pb,Davidson:2012ds}. As shown in
\cite{Harnik:2012pb}, in the case of final states including $\tau$
leptons, LHC data can already put constraints similar to those
from low-energy experiments. Therefore, the question is what type of
models allow for large Higgs LFV branching ratios and whether these
are compatible with the usual low-energy constraints.

This question was recently addressed in the context of R-parity
violating models in \cite{Arhrib:2012mg,Arhrib:2012ax}. The authors of
these references considered the MSSM extended to include all
renormalizable (L violating) RPV couplings and computed $\text{Br}(h
\to \mu \tau)$ at the 1-loop level. The key point in RPV models is the
particles-sparticles mixing induced by the RPV parameters and how it
leads to tree-level LFV Higgs decays.

Two examples are shown in figure \ref{fig:contributions}. On the
left-hand side, a $B \, \lambda$ contribution is shown, whereas on the
right-hand side a $\mu_i^2$ contribution is presented. Here $\mu_i$
and $\lambda_{ijk}$ are the superpotential couplings $\mathcal{W}
\supset \mu_i \widehat L_i \widehat H_u + \frac{1}{2} \lambda_{ijk}
\widehat L_i \widehat L_j \widehat{e}_k^c$ and $B_i$ is the soft SUSY
breaking parameter $\mathcal{L} \supset B_i \tilde L_i
H_u$ \footnote{We are following here the notation in
  \cite{Arhrib:2012mg,Arhrib:2012ax}. The $\mu_i$ parameters should
  not be confused with the usual Higgs superpotential mass term
  $\mu$.}.

\begin{figure}
\begin{center}
\vspace{5mm}
\includegraphics[width=0.48\textwidth]{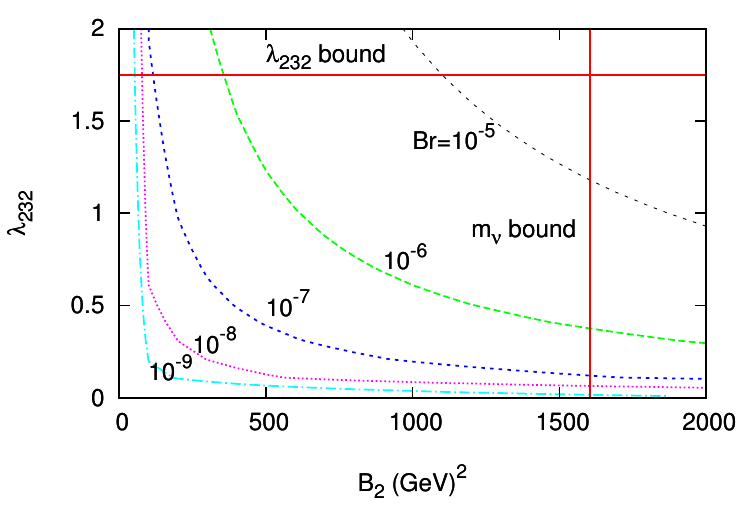}
\end{center}
\caption{$\text{Br}(h \to \mu \tau)$ contours on the $B_2 -
  \lambda_{232}$ plane. The red lines show approximate limits due to
  neutrino masses (in case of $B_2$) and charged current experiments
  (in case of $\lambda_{232}$). Figure borrowed from
  \cite{Arhrib:2012mg}.}
\label{fig:resulthmutau}
\end{figure}

The parameter combination that gives rise to the largest $\text{Br}(h
\to \mu \tau)$ is $B \lambda$, where the mixing between the Higgs
boson and the sneutrinos is combined with the lepton number violating
$\lambda$ coupling. However, the same combination of couplings
contributes to neutrino masses
\cite{Grossman:1998py,Chun:1999bq,Davidson:2000uc}. Moreover, the
$\lambda$ coupling is constrained by charged current experiments
\cite{Barbier:2004ez}. For these reasons, the largest value for
$\text{Br}(h \to \mu \tau)$ compatible with the previous bounds is
actually found to be quite small. This is shown in figure
\ref{fig:resulthmutau}, where several $\text{Br}(h \to \mu \tau)$
contours are drawn on the $B_2 - \lambda_{232}$ plane. From this
figure one can easily conclude that $\text{Br}(h \to \mu \tau)$ can
reach, at most, $\text{a few} \, \times \, 10^{-5}$.

Although this result is a little disappointing (the LHC sensitivity
with $20 \, fb^{-1}$ at $8$ TeV was estimated in \cite{Davidson:2012ds}
to be around $\text{Br} \sim 10^{-3}$), further investigation in
extended RPV scenarios might reveal better chances.

\subsection{Exotic muon decays}
\label{subsec:majoron}

A very interesting framework for lepton flavor violation is that of
spontaneous R-parity violation. In this case, and in order to
establish a bridge to neutrino masses, one considers a supersymmetric
theory that in principle conserves lepton number, but whose vacuum
breaks it\footnote{One could also think of spontaneous baryon number
  violation, but that is clearly out of the scope of this talk.}. If
the breaking is caused by the vacuum expectation value (VEV) of a scalar
field with an odd lepton number, R-parity is broken as well.

The main difference with respect to explicit R-parity violation is the
existence of a massless Goldstone boson, the Majoron ($J$)
\cite{Chikashige:1980ui,Gelmini:1980re}, associated to the spontaneous
breaking of the continuous $U(1)_L$ symmetry. The nature of this
massless state is actually of great importance for the
phenomenological success of the model. The first attemp in this
direction was made in \cite{Aulakh:1982yn}, where the breaking of
R-parity was caused by the VEV of a left-handed sneutrino. The model
was eventually excluded since the doublet nature of the majoron leads
to conflict with LEP bounds and astrophysical data
\cite{Raffelt:1996wa,Amsler:2008zzb}. However, more refined models,
where the violation of lepton number is induced by a gauge singlet,
are still valid possibilities. We will consider here the model
introduced in \cite{Masiero:1990uj}.

The model under consideration \cite{Masiero:1990uj} contains three
additional singlet superfields, namely, $\widehat\nu^c$, $\widehat S$
and $\widehat\Phi$, with lepton number assignments of $L=-1,1,0$
respectively. The superpotential can be written as
\begin{eqnarray} %
{\cal W} &=& Y_u^{ij} \widehat{Q}_i \widehat{u}_j^c \widehat{H}_u + Y_d^{ij} \widehat{Q}_i \widehat{d}_j \widehat{H}_d + Y_e^{ij} \widehat{L}_i \widehat{e}_j^c \widehat{H}_d %\nonumber
\\
        & + & Y_\nu^i\widehat L_i\widehat \nu^c\widehat H_u
          - h_0 \widehat H_d \widehat H_u \widehat\Phi
          + h \widehat\Phi \widehat\nu^c\widehat S +
          \frac{\lambda}{3!} \widehat\Phi^3 . \nonumber
\label{eq:Wsuppot}
\end{eqnarray}
For simplicity we consider only one generation of $\widehat\nu^c$ and
$\widehat S$. Adding more generations of $\widehat\nu^c$ or $\widehat
S$ does not add any qualitatively new features to the model. At low
energy, i.e.~after electroweak symmetry breaking, various fields
acquire VEVs. Besides the usual MSSM Higgs boson VEVs $v_d$ and $v_u$,
these are $\langle \Phi \rangle = v_{\Phi}/\sqrt{2}$, $\langle {\tilde
  \nu}^c \rangle = v_R/\sqrt{2}$, $\langle {\tilde S} \rangle =
v_S/\sqrt{2}$ and $\langle {\tilde \nu}_i \rangle = v_i/\sqrt{2}$.
Note, that $v_R \ne 0$ generates effective bilinear terms $\mu_i
= Y_\nu^i v_R/\sqrt{2}$ and that $v_R$, $v_S$ and $v_i$ violate lepton
number as well as R-parity.

Although the presence of a massless majoron is allowed by the
experimental contraints, it dramatically changes the phenomenology
both at collider and low-energy experiments
\cite{Hirsch:2008ur,Hirsch:2009ee}. In particular, it leads to new LFV
processes, such as $\mu \to e J$ or $\mu \to e J \gamma$. The exotic
decay $\mu \to e J$ was first studied in \cite{Romao:1991tp} and later
revisited in \cite{Hirsch:2009ee}, where the decay with an additional
photon was also considered.

The decays $\ell_i\to \ell_j J$ can be calculated from the general
coupling $\chi^+_i-\chi^-_j-P^0_k$. A straightforward computation
\cite{Hirsch:2009ee} shows that the $e-\mu-J$ coupling, $O_{e \mu J}$,
is of the form $\displaystyle O_{e \mu J} \sim \frac{1}{v_R} \, \times
\, \text{RPV parameters}$, which makes us conclude that, in general,
one expects large partial decay widths to majorons if $v_R$ is
low. However, searches for $\mu \to e J$ have to deal with a huge
irreducible background coming from the standard muon decay $\mu \to e
\nu \nu$. Moreover, there are no experiments looking for $\mu \to e J$
and the current limit on the branching ratio, $\text{Br}(\mu \to e J)
\lesssim 10^{-5}$, dates back to 1986
\cite{Jodidio:1986mz}\footnote{In fact, the bound we provide here is
  based on a reinterpretation of the results of \cite{Jodidio:1986mz},
  which considered a slightly different scenario.}. For these reasons,
we consider the decay $\mu \to e J \gamma$ which might be more
interesting due to the existent experiments looking for $\mu \to e
\gamma$.

It is easy to find a very simple relation between the two branching ratios
\begin{equation}
\text{Br}(\mu \to e J \gamma) = \frac{\alpha}{2 \pi} {\cal I}(x_{min},y_{min}) \text{Br}(\mu \to e J).
\end{equation}
Here ${\cal I}(x_{min},y_{min})$ is a phase space integral given by
\begin{equation}
{\cal I}(x_{min},y_{min}) = \int dx dy \frac{(x-1)(2-xy-y)}{y^2(1-x-y)},
\end{equation}
the dimensionless parameters $x$, $y$ are defined as usual
\begin{equation}
x = \frac{2 E_e}{m_\mu} \quad , \quad y = \frac{2 E_\gamma}{m_\mu}
\end{equation}
and $x_{min}$ and $y_{min}$ are the minimal electron and photon 
energies measured in a given experiment.

\begin{figure*}
\begin{center}
\vspace{5mm}
\includegraphics[width=0.4\textwidth]{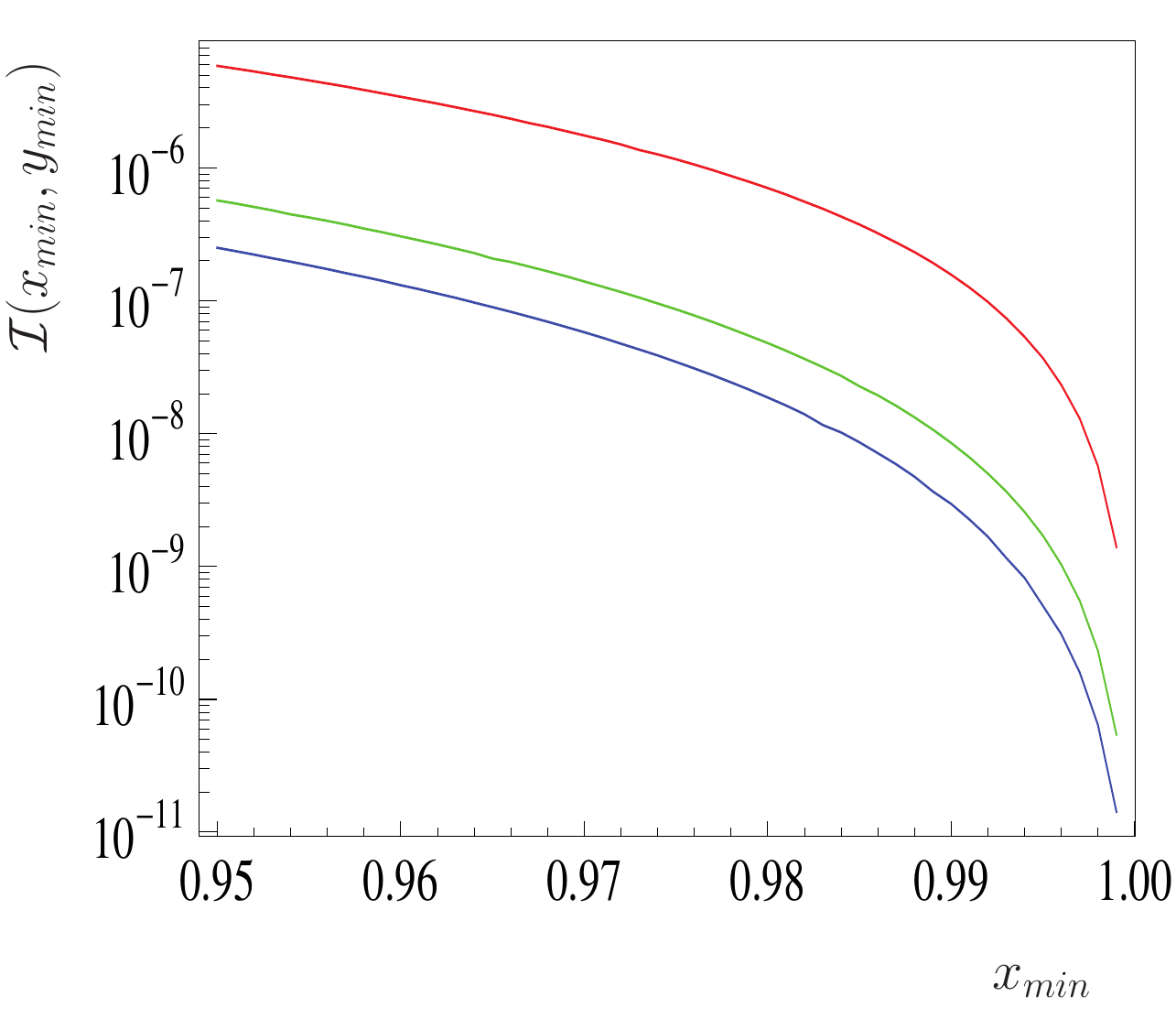}
\hspace*{5mm}
\includegraphics[width=0.4\textwidth]{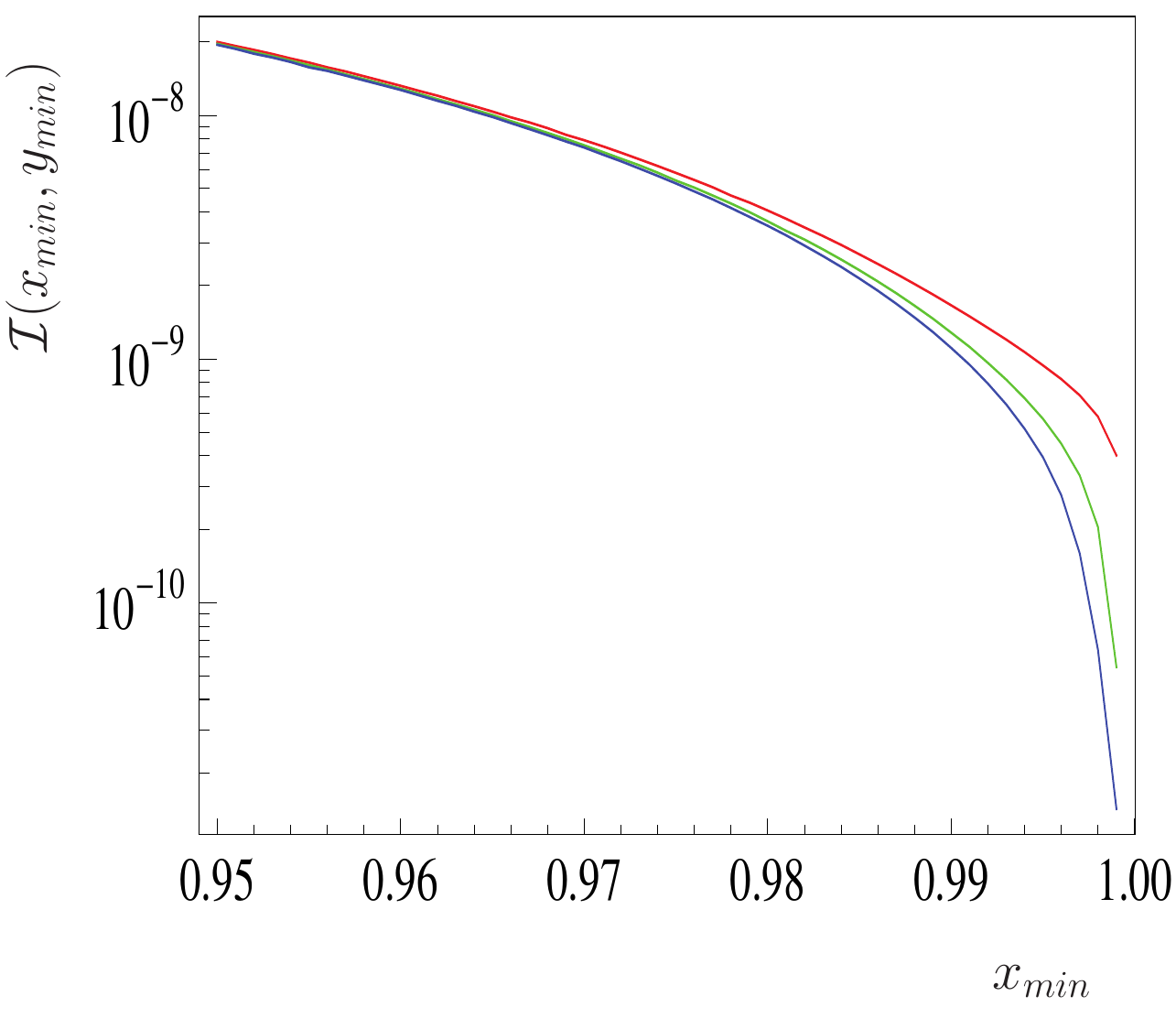}
\end{center}
\caption{The phase space integral for the decay $\mu\to e J \gamma$ 
as a function of $x_{min}$ for three different values of $y_{min}= 
0.95, 0.99, 0.995$ from top to bottom and for two different values of 
$\cos\theta_{e\gamma}$. To the left $\cos\theta_{e\gamma}=-0.99$, 
to the right $\cos\theta_{e\gamma}=-0.99997$.}
\label{fig:int}
\end{figure*}

The question now is whether the MEG experiment can look for this
process. Figure \ref{fig:int} shows the value of the phase space
integral ${\cal I}(x_{min},y_{min})$ as a function of $x_{min}$ for
three different values of $y_{min}$ and for two choices of
$\cos\theta_{e\gamma}$.  The MEG proposal describes the cuts used in
the search for $\mu\to e \gamma$ as $x_{min}\ge 0.995$, $y_{min}\ge
0.99 $ and $|\pi - \theta_{e\gamma}| \le$ 8.4 mrad. For these values
one finds ${\cal I} \simeq 6 \cdot 10^{-10}$. A limit for
$\text{Br}(\mu\to e \gamma)$ of $\text{Br}(\mu\to e \gamma)$ $\le
10^{-13}$ then translates into a limit of $\text{Br}(\mu\to e J) \le
0.14$, obviously meaningless. To improve upon this bound, it is
necessary to relax the cuts. For example, relaxing the cut on the
opening angle to $\cos\theta_{e\gamma}=-0.99$, the value of the
integral increases by more than 3 orders of magnitude for
$x_{min}=y_{min}\ge 0.95$.

On the other hand, such a change in the analysis is prone to induce
background events, which the MEG cuts were designed for to avoid. In
particular, accidental background from muon annihilation in
flight\footnote{Less important is the irreducible background induced
  by prompt events from the standard model radiative decay $\mu \to e
  \nu \nu \gamma$. Moreover, kinematical information might allow
  to discriminate this decay from $\mu\to e J \gamma$.}. Therefore,
although one could in principle increase the value of the phase space
integral ${\cal I}(x_{min},y_{min})$, the background in that case
would make the search for a positive signal impossible. Certainly, a
better timing resolution of the experiment would be required to reduce
this background.

\section{Supersymmetric low-scale seesaw models}
\label{sec:lowscale}

Let us now briefly discuss LFV in low-scale seesaw models. For more
information about the phenomenology in non-SUSY models see
\cite{hambyetalk}.

The most popular way to address neutrino masses is the famous seesaw
mechanism. In its traditional form, the seesaw mechanism explains the
smallness of neutrino masses thanks to the suppression by a high
scale, $\Lambda$, at which neutrino masses are generated. Several
realizations (supersymmetric or not) of this framework are known. In
the supersymmetric versions of these realizations, the low energy
spectrum is composed by the MSSM superfields, since the seesaw
messengers decouple at $\Lambda \gg m_{SUSY}$.

At the SUSY scale, the misalignment of the slepton mass matrices with
respect to those of the SM charged leptons induces lepton flavor
violation. However, different LFV processes may have very different
rates. The \emph{classical} supersymmetric seesaw usually predicts
that the branching ratios for the decays $\ell_i \to 3 \ell_j$ are roughly a
factor $\alpha$ smaller than those for the corresponding decays $\ell_i
\to \ell_j \gamma$ \cite{Ilakovac:1994kj,Hisano:1995cp,Arganda:2005ji},
\begin{equation} \label{eq:dipole}
\text{Br}(\ell_i \to 3 \ell_j) \simeq \frac{\alpha}{3 \pi} 
\left(\log\left(\frac{m^2_{\ell_i}}{m^2_{\ell_j}}\right) - \frac{11}{4} \right) 
\text{Br}(\ell_i \to \ell_j \gamma) \, .
\end{equation}
Therefore,
in these scenarios, which belong to the \emph{minimal SUSY} category,
the most relevant LFV process is $\ell_i \to \ell_j \gamma$.

The relation in Eq. \eqref{eq:dipole} can be traced back to the
so-called \emph{dipole dominance} in high-scale seesaw models. Among
all contributions to $\ell_i \to 3 \ell_j$, the dipole generated in photon
penguins is the dominant one. For this reason, the branching ratios
for $\ell_i \to \ell_j \gamma$ and $\ell_i \to 3 \ell_j$ turn out to be
proportional. However, the dipole dominance can be broken in
non-minimal scenarios that include new superfields besides those in
the MSSM. This is in fact the case of supersymmetric low-scale seesaw
models.

Contrary to the classical SUSY seesaw, in low-scale seesaw models the
new states (in addition to those of the MSSM) do not decouple: indeed,
the seesaw messengers can be as light as to be present at the SUSY
scale, leading to new contributions that can change the picture
dramatically~\footnote{A prime example of this class of models is the
  inverse seesaw \cite{Mohapatra:1986bd}. The inverse seesaw can be
  embedded in the MSSM by the addition of two extra gauge singlet
  superfields with opposite lepton numbers ($+1$ and $-1$). In this
  framework one can in principle have large neutrino Yukawa couplings
  compatible with a seesaw scale close to the SUSY one.}. This has
recently fuelled a significant number of studies. Let us now mention
some of them:

\begin{itemize}

\item {\bf Non-SUSY boxes:} Several authors have recently studied LFV
  in the presence of light right-handed neutrinos, paying special
  attention to $\ell_i \to 3\ell_j$ and $\mu - e$ conversion in nuclei
  \cite{Ilakovac:2009jf,Alonso:2012ji,Dinh:2012bp,Ilakovac:2012sh}. Surprisingly,
  it has been found that the main contributions to these two processes
  may come from (non-supersymmetric) box diagrams. These non-SUSY
  contributions tend to dominate for light right-handed
  neutrinos. Therefore, although they are not genuine SUSY
  contributions, they must be taken into account in any SUSY study.

\item {\bf Z-penguins:} The presence of light `right-handed'
  sneutrinos also gives rise to new contributions. In
  \cite{Hirsch:2012ax} it was pointed out that in this case the
  Z-penguins, usually regarded as subdominant, can in fact provide the
  dominant contributions to LFV processes such as $\ell_i \to 3\ell_j$
  and $\mu - e$ conversion in nuclei. This was further explored
  in~\cite{Dreiner:2012mx,Hirsch:2012kv,Abada:2012cq}, where
  enhancements to the abovementioned processes thanks to the
  supersymmetric Z-penguins were found in several contexts.

\end{itemize}

In both cases, the relation in Eq. \eqref{eq:dipole} is not fulfilled
and one can actually have $\ell_i \to 3\ell_j$ and $\mu - e$
conversion in nuclei as the most constraining observables.

\section{Final remarks}
\label{sec:final}

In this talk we have reviewed some recent developments in lepton
flavor violation in non-minimal SUSY models. The bottom line is that
lepton flavor violation may be much richer than expected when one goes
to non-minimal supersymmetric extensions of the leptonic sector. In most
models the common lore (established in the MSSM) turns out to be
completely wrong, and thus specific studies must be performed in order
to correctly describe the corresponding phenomenology. This in fact
translates into two messages: (1) for the theorists, lepton flavor
violation might be much more intricate than what minimal models
predict. We should be open-minded and consider non-standard scenarios
where the leptonic sector is extended beyond the MSSM realization. And
(2) for the experimentalists, although minimal models are well
motivated, lepton flavor violation might show up in non-standard
channels. We must be ready to find a signal there as well.

\nocite{*}
\bibliographystyle{elsarticle-num}
\bibliography{refs_a_vicente}

\end{document}